\documentclass[showpacs,prb]{revtex4}
\usepackage{epsfig}
\newcommand {\nulla}[1]{}

\newcommand {\nb}{\nonumber\\}

\newcommand {\be}{\begin{equation}}
\newcommand {\ee}{\end{equation}}
\newcommand {\bea}{\begin{eqnarray}}
\newcommand {\ea}{\end{eqnarray*}}
\newcommand {\ba}{\begin{eqnarray*}}
\newcommand {\eea}{\end{eqnarray}}

\newcommand {\ket}{\rangle}

\newcommand {\rv}{{\bf r}}

\newcommand {\xv}{{\bf x}}

\newcommand {\yv}{{\bf y}}

\newcommand{\bem}{\begin{multline}}

\newcommand{\bit}{\begin{itemize}}
\newcommand{\eit}{\end{itemize}}

\newcommand{\ic}{\begin{center}}
\newcommand{\fc}{\end{center}}

\begin{document}
\title{ Microscopic study of the He$_2$-SF$_6$ trimers}
\author{P. Barletta, A. Fabrocini, A. Kievsky}
\affiliation{
Dipartimento di Fisica "E.Fermi", Universit\`a di Pisa, and  
INFN, Sezione di Pisa, Via Buonarroti, 2, I-56100 Pisa, Italy}
\author{J. Navarro}
\affiliation{IFIC (CSIC-Universidad de Valencia), Edificio 
Institutos de Paterna,
Apartado Postal 22085, E-46.071-Valencia, Spain} 
\author{A. Polls}
\affiliation{Departament d'Estructura i Constituents de la Mat\`eria,
Facultat de F\'{\i}sica, Universitat de Barcelona, 
E-08028, Barcelona, Spain}

\date{\today}


\begin{abstract}
The He$_2$-SF$_6$ trimers, in their different He isotopic combinations, are 
studied both in the framework of the correlated Jastrow approach and 
 of the Correlated Hyperspherical Harmonics expansion method. 
The energetics and structure of the He-SF$_6$ dimers are analyzed,
and the existence of a characteristic  rotational band in the excitation 
spectrum is discussed, as well as the isotopic differences. The binding 
energies and the spatial properties of the trimers, in their ground and 
lowest lying excited states, obtained by the Jastrow 
ansatz are in excellent agreement with the results of the converged 
CHH expansion. 
The introduction of the He-He correlation  makes all trimers bound by  
largely suppressing the short range He-He repulsion.   
 The structural properties of the trimers are qualitatively explained in 
terms of the shape of the interactions, Pauli principle and masses of 
the constituents. 
\end{abstract}


\maketitle

\section{Introduction}
Helium systems are dominated by quantum effects and remain liquid down to
zero temperature. This is a consequence of both the small atomic mass and the 
weak atom-atom interaction, which is the weakest among the rare gas atoms. 
Helium clusters remain liquid under all conditions of formation and are very
weakly bound systems.

Small helium clusters have been detected by diffraction of a helium nozzle
beam by a transmission grating.\cite{Sch94} Using a grating of 200 nm period, 
conclusive evidence of the existence of the dimer $^4$He$_2$ has been 
established.\cite{Sch96} The existence of $^4$He$_2$ was previously 
reported\cite{Luo93} using electron impact ionization techniques. Diffraction
experiments from a 100 nm period grating has lead to the determination of
a molecular bond length of $54\pm4$~\AA, out of which a binding energy of 
$1.1+0.3-0.2$~mK has been deduced.\cite{Gris01} This energy is in good
agreement with the values obtained by direct integration of the Schr\"odinger
equation using modern He-He interactions.\cite{Jan95} Theoretical calculations 
indicate that any number of $^4$He atoms form a self-bound system. In contrast,
a substantially larger number of $^3$He atoms is necessary for self-binding,
as a consequence of  the smaller $^3$He atomic mass and its fermionic nature.
The required minimum number has been estimated to be 29 atoms, using a density
functional plus configuration interaction techniques to solve the many-body 
problem,\cite{Bar97} or 34-35 atoms, using an accurate variational wave
function\cite{Gua00,Gua00a} with the HFD-B(HE) Aziz interaction.\cite{Azi87} 
It is worth recalling that a theoretical description of pure $^3$He clusters, 
either based on ab initio calculations or employing Green Function or
Diffusion Monte Carlo techniques, is still missing.

Doping a helium cluster with atomic or molecular impurities constitutes a
useful probe of the structural and energetic properties of the cluster itself.
It has been proved that rare gases and closed-shell molecules as HF, OCS
or SF$_6$ are located in the bulk of the cluster.~\cite{interior} Doped $^4$He 
clusters have been extensively studied in the past by a variety of methods, 
ranging from Diffusion and Path Integral Quantum Monte Carlo methods to 
two-fluid models. A comprehensive view of this subject is found in 
Ref.~\onlinecite{Kwo00}, giving account on a microscopic basis of the free 
rotation of a heavy molecule in a $^4$He nanodroplet, consistently with the 
occurring bosonic superfluidity at the attained temperatures. This phenomenon 
is not expected to take place in doped $^3$He clusters, because of the 
fermionic nature of the atoms, unless very low temperatures (below a few mK) 
are  reached. Even more interesting, from both the theoretical and experimental 
points of view, are mixed doped clusters of $^3$He and $^4$He atoms. The status 
of the theory in these last cases is far behind that in the $^4$He droplets.
The most updated studies of $^3$He and mixed $^4$He-$^3$He clusters, either
pure or doped, employ a finite-range density functional
theory.\cite{Bar97b,Gar98} 

In this work we study the properties of a trimer formed by two helium atoms 
plus a heavy dopant. The dopant molecule behaves as an attractive center binding
a certain number of otherwise unbound $^3$He atoms. This fact has been used in 
Ref.~\onlinecite{Jun01} to set an analogy between electrons bound by an atomic
nucleus and $^3$He atoms bound by a dopant species. Systems formed by two $^3$He
atoms plus a molecule have been studied employing the usual quantum chemistry 
machinery. 

Helium droplets doped with the SF$_6$ molecule have been widely investigated, 
also in view of the fact that the interaction He-SF$_6$ is well 
established.\cite{Pac84,Tay93} In this work we use the spherically averaged 
interaction of Taylor and Hurly\cite{Tay93} between the helium atom and the 
SF$_6$ molecule, and the Aziz HFD-B(HE) helium-helium interaction.\cite{Azi87} 
We first perform a variational study of the $^3$He$_2$-SF$_6$, $^4$He$_2$-SF$_6$
and $^4$He-$^3$He-SF$_6$ trimers using a Jastrow correlated wave function. 
Then the Correlated Hyperspherical Harmonics (CHH) expansion method\cite{CHH} 
is employed and its outcomes are used as benchmarks for the variational 
calculations. The CHH expansion has shown to be a powerful technique to study 
three-and four-body strongly interacting systems. In light atomic nuclei its 
accuracy is comparable with (and in some cases even better than) other popular 
approaches, as the Faddeev, Faddeev-Yakubowsky and Quantum Monte Carlo 
ones.\cite{CHH1} Besides accurately studying the ground and first excited 
states of the trimers, comparing the variational and CHH results may provide 
essential clues for the construction of a reliable variational wave function 
to be used in heavier doped nanodroplets.

The plan of this paper is the following. In Sect. II we study the dimers formed 
by a single helium atom and the SF$_6$ molecule and enlighten some aspects of 
their excitation spectrum. In Sect. III we consider the trimers by the Jastrow 
variational and CHH approaches. Results for the energetics and structure of the 
trimers are given and discussed in Section IV. Finally, Sect. V provides the 
conclusions and the future perspectives of this work.

\section{The He-SF$_6$ dimer}
Prior to the study of trimers made by two helium atoms and a dopant species,
it is convenient to analize the dimer in some details. To fix the notation for 
later discussions, we write here the Schr\"odinger equation for the relative 
motion of a helium atom and a dopant, $D$,
\begin{equation}
h_0 \phi_{n\ell}({\bf r}) = e_{n\ell} \phi_{n\ell}({\bf r})
\label{schroedinger_0}
\end{equation}
with
\begin{equation}
h_0 = - \, \frac{\hbar^2}{2m_{\alpha}} \nabla^2 + V_{\rm He-D}({\bf r})
\label{schroedinger_00}
\end{equation}
where $m_{\alpha}$ (with $\alpha=3,4$) is the reduced mass of the 
$^\alpha$He-D pair, and
$V_{\rm He-D}({\bf r})$ is the helium-dopant  interaction, being ${\bf r}$ the
relative coordinate. We have numerically solved this equation for the dopant
SF$_6$, using the spherically averaged interaction determined in 
Ref.~\onlinecite{Tay93}. A set of energies and orbitals characterized by the 
quantum numbers $(n\ell)$ is thus obtained. 

The He-SF$_6$ interaction has an attractive well strong enough to sustain 12 
and 15 bound states for isotopes $^3$He and $^4$He, respectively. In 
Table~\ref{single_obser} some observables of the dimers are displayed. The 
calculations have been done in the limit of infinite mass of the SF$_6$ 
molecule. The most striking feature of the energy spectra is that the first  
7 (9) levels correspond to nodeless $n=0$ states. 
Notice that for each isotope the expectation values 
$\langle R \rangle$ and $\sqrt{\langle R^2 \rangle}$ are not very different, 
neither for a given $\ell$-state nor for different $\ell$ values. The 
expectation value $\langle V \rangle$ neither varies too much,  
increasing by $\sim 10\%$ in going from the $\ell$=0 to the 
$\ell$=8,9 states. 
These results are an indication that the wave functions show a 
well-defined peak in nearly the same region, as a consequence of the 
characteristics of the He-SF$_6$ interaction. 

To first order in the mass ratio, the correction to the infinite 
dopant mass approximation modifies the kinetic energy as 
$T^{(\alpha)}=T^{(\alpha)}_{M=\infty}(1+m_\alpha/M)$,
 with $M$ the dopant mass. 
Accordingly, the finite mass system results
 less bound by $\Delta_E^{(\alpha)}=T^{(\alpha)}_{M=\infty}(m_\alpha/M)$. 
The mass ratios $m_\alpha/M$ are $0.0207$ and $0.0274$ for $^3$He
 and $^4$He, respectively, considering the $^{32}$S isotope. 
 So, the binding energy corrections are 
 $\Delta_E^{(3)}=0.245$~K and $\Delta_E^{(4)}=0.300$~K, both being 
about 1$\%$ of the total energy. An exact finite mass calculation 
confirms this estimate,  
providing $\Delta_E^{(3)}=0.242$~K and $\Delta_E^{(4)}=0.296$~K.

\begin{figure}
 \includegraphics*[angle=-90, width=15cm]{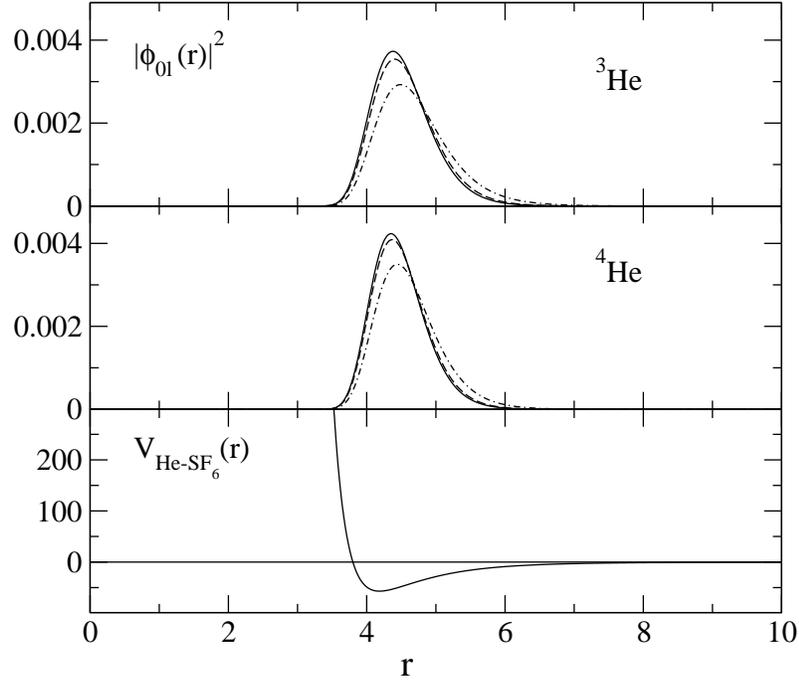}
\caption{Dimer radial probability densities 
($^3$He-SF$_6$: upper panel; $^4$He-SF$_6$: middle panel); 
He-SF$_6$ interaction (lower panel). 
The probability densities correspond to the quantum
numbers $\ell=0$ (solid line), $\ell=4$ (dashed line) and
$\ell=8$ and 9 (dash-dotted line) for $^3$He and 
$^4$He, respectively.
Distances in \AA, potential in K and densities in \AA$^{-3}$. 
}
\label{probdens}
\end{figure}

\begin{figure}
\includegraphics*[width=12cm]{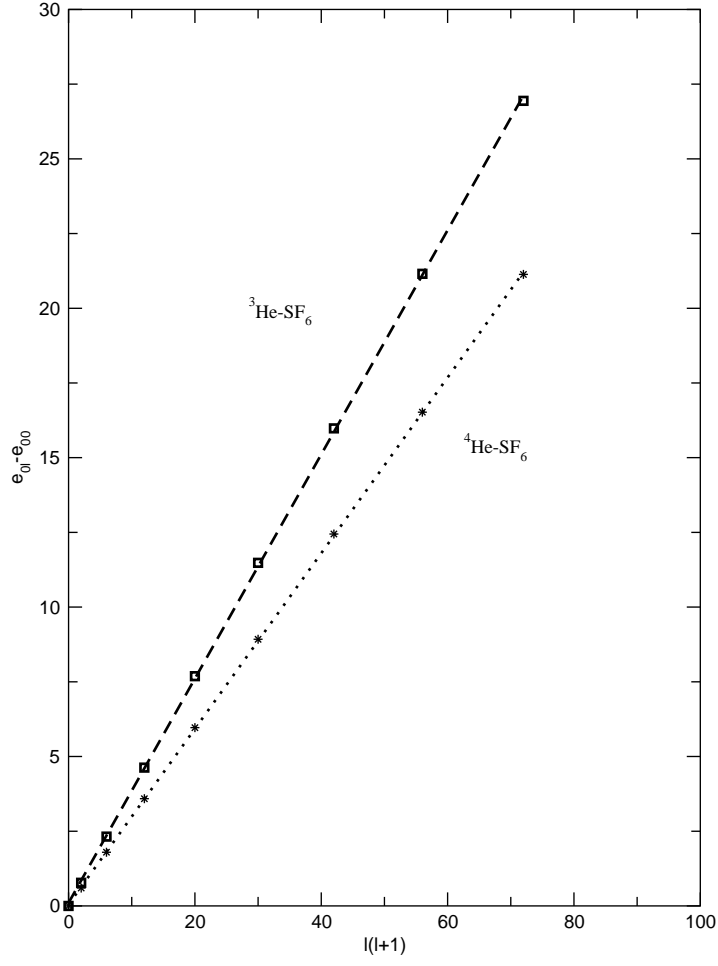}
\caption{Differences $e_{0l}-e_{00}$ (in K) vs. $\ell(\ell+1)$ for the 
the He-SF$_6$ dimers. Squares correspond to the $^3$He-SF$_6$ results and 
stars to the $^4$He-SF$_6$ ones. The lines are linear fits to the 
numerical values.}
\label{rotational}
\end{figure}

The He-SF$_6$ interaction is displayed in Fig.~\ref{probdens}. It is strongly 
repulsive at distances shorter than $\simeq3.8$~\AA, and attractive beyond. 
The attractive part is mostly concentrated in a narrow region  around 4.2~\AA, 
immediately after the repulsive core. This implies that the He atom locates 
relatively far away from the potential origin, so that the centrifugal term 
entering the Schr\"odinger equation can be considered as a perturbation. Two 
consequences can be deduced from this observation. First, the radial 
distributions are very peaked in the same narrow region, independently on the 
value of the angular momentum $\ell$, as it is shown in Fig.~\ref{probdens} 
for three states, corresponding to $\ell=0$ and 4 for both isotopes, and the 
nodeless bound states with the highest excitation energy, namely $\ell=8$ for 
$^3$He and 9 for $^4$He. The distributions are concentrated in the same region, 
and those corresponding to $\ell=0$ and 4 are barely distinguishible. 
Second, the excitation energies are closely proportional to $\ell (\ell+1)$, 
i.e. they follow a rotational pattern. Figure \ref{rotational} depicts 
the differences $(e_{0\ell}-e_{00})$ in functions of $\ell(\ell+1)$  
(squares for $^3$He-SF$_6$ and stars for $^4$He-SF$_6$), together with the  
linear fits to the energy differences. The slopes provide the $n=0$ 
rotational constants, $C_0^{(3,4)}$. From the the fits we obtain 
$C_0^{(3)}=0.376$~K and $C_0^{(4)}=0.294$~K, in good agreement with 
the rough  estimate $\hbar^2/2m\langle R^2\rangle$, 
where $\langle R^2\rangle$ is the mean square Dopant-Helium distance. 
A similar behavior is found for the $n=1$ excited states, whose  rotational 
constants are $C_1^{(3)}=0.194$~K and $C_1^{(4)}=0.179$~K. 
The ratio of the rotational constants of the dimers with either isotope 
are of course in the inverse ratio of masses.  Notice that 
the larger mass of the $^4$He atom translates into a larger binding. Both the 
decrease of the kinetic energy and the increase of the attraction contribute 
to the increment of the binding energy for the $^4$He-SF$_6$ dimer. The stronger
localization of the $^4$He atom is visualized by the peak of the radial 
probability density shown in Fig.~\ref{probdens}, which is slightly higher for 
the $^4$He atom.

It is worth stressing that the observed rotational spectrum is a direct 
consequence of the shape of the  He-SF$_6$ interaction, whose attractive well, 
having a depth of $\approx-57$~K, allows for twelve or fifteen bound states. 
Taking the lighter Ne as a dopant, we have found that only three bound states 
exist, being the depth of the attractive well $\approx-20$~K. Moreover, the 
binding energies are very small, and the probability distributions are extended 
over a large region.   

\section{The He$_2$-SF$_6$ trimers: theory}
In the limit of infinite mass of the dopant molecule, hence considered as a 
fixed center, the Hamiltonian of the  He$_2$-SF$_6$ trimers is
\begin{equation}
H(1,2) = -
\sum_{i=1}^2 {\hbar^2 \over 2m_{\alpha_i}} \nabla^2_i + 
\sum_{i=1}^2 V_{\rm D-He}({\bf r}_i) + V_{\rm He-He}({\bf r}_{12})  \,
\label{hamiltonian}
\end{equation}
with $\alpha_i=3,4$.  $V_{\rm D-He}({\bf r})$ and $V_{\rm He-He}({\bf r})$ 
are the dopant--helium and helium--helium interaction potentials, respectively. 
${\bf r}_{i}$ is the coordinate of the $i$--th helium atom with respect to 
the central molecule and ${\bf r}_{12}$ is the helium--helium relative 
coordinate.

The ground and excited states properties of the trimers can be obtained either 
 by an exact solution of the Schr\"odinger equation,
\begin{equation}
H(1,2) \, \Psi_{\gamma}^T(1,2) \,  = \, E_{\gamma} \, \Psi_{\gamma}^T(1,2) \, 
\label{schroedinger}
\end{equation}
or by some approximate estimate of their wave functions. Here $\gamma$ labels 
the generic trimer state, whose wave function is $\Psi_{\gamma}^T(1,2)$.  

The Schr\"odinger equation for clusters of $^4$He atoms may be exactly solved 
for the ground state by quantum Monte Carlo (QMC) 
methods.\cite{Chi92,Bar93,Lew97} Other approaches, as Variational Monte 
Carlo\cite{Pan86,Chi95,Gua01} (VMC) with Jastrow correlated wave functions or 
density functional theories\cite{Str87,Dal95} (DFT), provide a less accurate 
description of the clusters. However, they are generally more flexible than QMC.
 
The presence of more than two $^3$He atoms makes the exact solution of 
the Schr\"odinger equation
 much more difficult, because of the notorious sign 
problem\cite{sign} associated to their fermionic nature. As a consequence, 
only DFT based studies of doped $^3$He clusters are available in 
literature.\cite{Bar97b,Gar98} The most updated study of the $^3$He$_2$-SF$_6$ 
trimer has been done within the Hartree-Fock (HF) approximation, not considering
the strong He-He correlations induced by $V_{\rm He-He}$.\cite{Jun01} 

In this Section we will first present a variational approach based on a 
Jastrow (J) correlated wave function. Then we will apply the Correlated 
Harmonical Hyperspherical (CHH) expansion method\cite{CHH} to further 
improve the description of the doped trimers.

\subsection{Variational approach}

The Jastrow correlated wave function of the trimer for the $\gamma$--state 
is given by:
\begin{equation}
\Psi_{\gamma}^J(1,2) \,  = \, \Phi_{\gamma}(1,2) \, f_J(r_{12}) \,
\label{correlated}
\end{equation}
where $\Phi_{\gamma}(1,2)$ is an independent particle (IP) wave function of 
the two helium atoms in the dopant field having the same set, $\gamma$, of 
quantum numbers. The correlation function between the two atoms, $f_J(r)$, 
is assumed to depend only on the interatomic distance and takes into account 
the modification to the IP wave function mainly due to the 
He-He  interaction. The optimal 
$f_J(r)$ is variationally fixed by minimizing the total energy of the state. 
$\Phi_{\gamma}(1,2)$ is built as an appropriate combination of the dimer 
He-SF$_6$ wave functions. For instance, using the $(LS)^\pi$ notation  
($L$ being the IP orbital angular momentum, $S$ the total spin and $\pi$ 
the parity of the helium pair), the $(00)^+$ IP wave function for 
the $^3$He$_2$-SF$_6$ trimer is taken as:
\begin{equation}
\Phi_{(00)^+}(1,2) \,  = 
\, \Xi_{0}(1,2) \, \phi_{1s}({\bf r}_1)\phi_{1s}({\bf r}_2) \, 
\label{1s1s}
\end{equation}
where $\Xi_{0}(1,2)$ is the spin-singlet wave function of the $^3$He-$^3$He 
pair and $\phi_{1s}({\bf r})$ is the $1s$ ($n$=0 and $\ell$=0) solution 
of the $^3$He-SF$_6$ dimer Schr\"odinger equation (\ref{schroedinger_0}).

In an analogous way, the $(10)^-$ and  $(11)^-$ IP wave functions are:
\begin{equation}
\Phi_{(10)^-}(1,2) \,  = \, \Xi_{0}(1,2) \, {1 \over {\sqrt 2}} 
[ \phi_{1s}({\bf r}_1)\phi_{1p}({\bf r}_2) + 
\phi_{1p}({\bf r}_1)\phi_{1s}({\bf r}_2)] \, 
\label{1s1p+}
\end{equation}
and
\begin{equation}
\Phi_{(11)^-}(1,2) \,  = \, \Xi_{1}(1,2) \, {1 \over {\sqrt 2}} 
[ \phi_{1s}({\bf r}_1)\phi_{1p}({\bf r}_2) - 
\phi_{1p}({\bf r}_1)\phi_{1s}({\bf r}_2)] \, 
\label{1s1p-}
\end{equation}
where $\Xi_{1}(1,2)$ is the spin-triplet pair wave function.

The total Hamiltonian (\ref{hamiltonian}) can be written as:
\begin{equation}
H(1,2) = h_0(1)+h_0(2) + V_{He-He}({\bf r}_{12}) = H_0(1,2) + V_{He-He}({\bf r}_{12})  \, 
\label{hamiltonian_1}
\end{equation}
where $h_0$ is the Hamiltonian (\ref{schroedinger_00}) of the dimer, and
\begin{equation}
H_0(1,2) \, \Phi_{\gamma }(1,2) \,  = \, 
(e_{\gamma_1} + e_{\gamma_2} ) \, \Phi_{\gamma}(1,2) \, 
\label{schroedinger_1}
\end{equation}
where the two sets of dimer quantum numbers, $\gamma_{1}$ and $\gamma_{2}$, 
are those taken to build up the total $\gamma$-trimer state. 

The total energy of the trimer in the  $\gamma$-state is 
\begin{equation}
E_{\gamma}={
{\langle \Psi^J_\gamma \vert H(1,2) \vert \Psi^J_\gamma \rangle} 
\over
{\langle \Psi^J_\gamma \vert \Psi^J_\gamma \rangle}
} \, .
\label{energy}
\end{equation}
Since $[V_{D-He},f_J]=0$, the energy results to be:
\begin{equation}
E_{\gamma}=e_{\gamma_1}+e_{\gamma_2} + 
\sum_{i=1}^2 {\hbar^2 \over 2m_{\alpha_i}} {
{\langle \Phi_\gamma \vert  
\nabla_i f_J(r_{12}) \cdot \nabla_i f_J(r_{12}) 
 \vert \Phi_\gamma \rangle} 
\over
{\langle \Psi^J_\gamma \vert \Psi^J_\gamma \rangle}
} +  {
{\langle \Phi_\gamma \vert  
 f_J(r_{12})  V_{He-He}({\bf r}_{12})  f_J(r_{12}) 
 \vert \Phi_\gamma \rangle} 
\over
{\langle \Psi^J_\gamma \vert \Psi^J_\gamma \rangle}
} \, . 
\label{energy_J}
\end{equation}
This equation will be used to estimate the variational energy 
of the trimer and to optimize the choice of the correlation factor.

\subsection{CHH approach}

In order to implement the CHH method for 
a system of three atoms of masses $m_i$, in positions $\rv_i$, it
is convenient to introduce the  
three sets $(\xv_i,\yv_i)$ of Jacobi coordinates: 
\bea
\begin{array}{l}
\xv_i=\sqrt{\frac{m_k m_j}{m_j+m_k}}(\rv_j-\rv_k) \, , \\
\yv_i=\sqrt{\frac{m_i (m_j+m_k)}{m_i+m_j+m_k}}
(\rv_i-\frac{m_j\rv_j+m_k\rv_k}{m_j+m_k}) \, .
\end{array}
\eea
In the fixed-center limit ($m_3=\infty$) the position $\rv_3$ coincides with 
the molecular center of mass and, for two equal mass 
atoms ($m_1=m_2=m$), the Jacobi coordinates, after dividing by 
$m^{1/2}$, can be reduced to:
\be
\left\{
\begin{array}{l}
\xv_1=\rv_2  \\
\yv_1=\rv_1 
\end{array}
\right. ,\,
\left\{
\begin{array}{l}
\xv_2=-\rv_1 \\
\yv_2=\rv_2 
\end{array}
\right. ,\,
\left\{
\begin{array}{l}
\xv_3= \sqrt{\frac{1}{2}}(\rv_1-\rv_2) \\
\yv_3=-\sqrt{\frac{1}{2}}(\rv_1+\rv_2)
\end{array}
\right. .
\ee

The total wave function, $\Psi_{(LS)^\pi}(1,2)$, can be expressed as 
a sum of three Faddeev-like amplitudes,\cite{CHH1} 
each of wich explicitely depends upon a different Jacobi set:
\be
\Psi_{(LS)^\pi}(1,2) = 
\psi^{(1)}_{(LS)^\pi} (\xv_1,\yv_1) + 
\psi^{(2)}_{(LS)^\pi} (\xv_2,\yv_2) + 
\psi^{(3)}_{(LS)^\pi} (\xv_3,\yv_3) \, .
 \label{psiexp}
\ee
The amplitudes are then expanded into channels, 
labelled by the partial angular momenta, $\ell_{x,i}$ 
and $\ell_{y,i}$, associated with $\xv_i$ and $\yv_i$, respectively:
\be
\psi^{(i)}_{(LS)^\pi}(\xv_i,\yv_i) =
\Xi_{S}(1,2) \sum_{\ell_{x,i},\ell_{y,i}}
\Phi^{(i)}_{\ell_{x,i},\ell_{y,i}}(x_i,y_i)
 [Y_{\ell_{x,i}}(\hat x_i) Y_{\ell_{y,i}}(\hat y_i)]_{LM}
 \, ,
\ee
where $Y_\ell$ are ordinary spherical harmonics, and $\Phi^{(i)}_{\ell_{x,i},\ell_{y,i}}$ is a two-dimensional 
function depending upon the moduli of the Jacobi vectors. 
As a result, the parity of the state is given by 
$\ell_{x,i}+\ell_{y,i}$,
 even (odd) for positive (negative) parity.

Following Ref.~\onlinecite{bk2001}, each amplitude 
 $\Phi^{(i)}_{\ell_{x,i},\ell_{y,i}}$ is expanded in terms of 
a correlated hyperspherical harmonics basis set. 
After introducing the hyperspherical coordinates, 
$(\rho,\phi_i)$, associated with the Jacobi set:
\bea
\rho^2 &=& x_1^2+y_1^2=x_2^2+y_2^2=x_3^2+y_3^2 \, , \\
x_i&=&\rho \cos\phi_i \, , \\
y_i&=&\rho \sin\phi_i \, , 
\eea
the CHH basis elements, having quantum numbers $(LS)^\pi$ 
and corresponding to the set of Jacobi coordinates labelled by $i$, 
are defined as follows: 
\bea
|m,k,\ell_{x,i},\ell_{y,i} ; i \ket &=& 
\Xi_{S}(1,2) F_J(r_1,r_2,r_{12}) 
\rho^{\ell_{x,i}+\ell_{y,i}} 
L^{(5)}_m(z)\exp(-{z \over 2}) \nonumber \\ 
&\times&  ^{(2)}P^{\ell_{x,i},\ell_{y,i}}_k(\phi_i) 
[Y_{\ell_{x,i}}({\hat x}_i)Y_{\ell_{y,i}}({\hat y}_i)]_{LM} \, , 
\label{chh}
\eea
where 
$^{(2)}P^{\ell_x,\ell_y}_k(\phi)=N_{\ell_x,\ell_y,k}
(\cos\phi)^{\ell_x}\,(\sin\phi)^{\ell_y} 
 P^{\ell_y+1/2,\ell_x+1/2}_k(\cos(2\phi))$
is a normalized hypherspherical polynomial, $P^{\ell_y+1/2,\ell_x+1/2}_k$ 
is a Jacobi polynomial, 
$L^{(5)}_m$ are associated Laguerre polynomials,  
 and $z=\beta\rho$, with $\beta$ a non-linear variational parameter. 

The other ingredient in the CHH basis elements is the correlation factor,  
$F_J (r_1,r_2,r_{12})$. 
Due to the strongly repulsive core of the interatomic potentials, 
it is convenient to take $F_J$ as a product of pairwise Jastrow-like 
correlation functions,
\be
F_J(r_1,r_2,r_{12})=g(r_1)g(r_2)f_J(r_{12}) \, .
\label{F_J}
\ee

The correlation functions mainly describe the short-range behavior 
of the wave function as the two helium atoms are close to each other 
or to the dopant. The polynomial part of the CHH expansion is expected 
to reproduce the mid- and long-range configurations. 
Therefore, it is preferable to choose correlation functions which 
behave in the outer portion of the Hilbert space as smoothly as possible, 
in order to avoid potentially deleterious biasing of the polynomial expansion. 

Suitable correlation functions to be used in (\ref{F_J}) 
are obtained by solving the two-body Schr\"odinger equation:
\be
\left(-\frac{\hbar^2}{2\mu}\nabla^2+V^*(r)
+\lambda e^{-\xi r}\right)h(r)=0 \, ,
\label{correq}
\ee
where $\mu$ is the reduced mass of the considered two-body system, 
and the pseudo-potential, $\lambda e^{-\xi r}$, is introduced 
to adjust the asymptotic behavior of the correlation function in 
such a way that $h(r\rightarrow\infty)\rightarrow 1$. 
The parameters $\xi$ and $\lambda$ are optimized for each 
of the different cases, SF$_6$-He, $^4$He-$^4$He and $^3$He-$^3$He.
For the helium-dopant correlation ($h=g$), 
we take $\mu=m_{\rm He}$, and $V^*$ is a modified SF$_6$-He potential. 
In fact, in order to build a nodeless correlation function, it is 
necessary to reduce the attractive part of the helium-dopant interaction. 
The repulsive part, on the other hand, is kept unaltered. 
In the $^4$He-$^4$He case ($h=f_J$, $\mu=m_{^4He}/2$), 
the correlation function has been obtained as in Ref.~\onlinecite{bk2001}. 
The $^3$He-$^3$He pair does not support a bound state, so the 
correlation function has been obtained simply as the solution 
of the zero-energy Schr\"odinger equation ($V^*=V_{He-He}$ 
and $\lambda=0$ in Eq.~\ref{correq}). 

In order to work with basis elements of defined symmetry under the permutation operator $\Pi_{12}$ (according to the Pauli principe) we take proper symmetric or antisymmetric combinations of the basis elements with 
$i=1$ and $2$:
\be
|m,k,\ell_{x,1},\ell_{y,1} ; \nu \ket = |m,k,\ell_{x,1},\ell_{y,1} ;1\ket + (-1)^q|m,k,\ell_{x,1},\ell_{y,1} ;2\ket  \, ,
\ee
where $\nu$ $(\equiv s,a)$ labels symmetric or antisymmetric states, 
respectively, and $q$=0, 1 is choosen according to the values of $\ell_x$ 
and $S$. 

The expansion of the total wave function in terms of CHH states with 
well defined permutation symmetry results in:  
\be
\Psi_{(LS)^\pi} = \sum_{\ell_{x,1},\ell_{y,1}}\sum_{k_a,m_a} A_{k_a,m_a} |k_a,m_a,\ell_{x,1},\ell_{y,1}; \nu\ket 
 + \sum_{\ell_{x,3},\ell_{y,3}}\sum_{k_b,m_b} B_{k_b,m_b} |k_b,m_b,\ell_{x,3},\ell_{y,3}; 3\ket \, .
\label{chhex}
\ee

The sum over the partial angular momenta $\ell_x,\ell_y$, although 
constrained by the values of the total angular momentum $L$, the parity 
and by  symmetry considerations, runs over an infinite number of channels. 
However, in practice only the lower channels are included, since 
the higher the angular momentum the less the contribution to the wave function. 
Moreover, due to the presence of both the correlation factors $F_J$ 
and the amplitude expansion, an infinite number of channels is 
automatically included, though in a non-flexible way. 

In the description of the $(LS)^\pi=(00)^+,(10)^-,(11)^-$ states 
for $^3$He$_2$-SF$_6$, we have retained only the lowest angular 
momentum channels, that is, $\ell_{x,1}=\ell_{y,1}=\ell_{x,3}=\ell_{y,3}=0$ 
for $(00)^+$, $\ell_{x,1}=\ell_{x,3}=0$ and $\ell_{y,1}=\ell_{y,3}=1$  
for $(10)^-$, $\ell_{x,1}=\ell_{x,3}=1$ and $\ell_{y,1}=\ell_{y,3}=0$ 
for $(00)^-$. 
The CHH wave functions for these three states are:
\bea
&&\Psi_{(00)^+}(1,2) = 
\Xi_{0}(1,2) F_J e^{-z/2} 
\left\{ \sum_{k_a,m_a} A_{k_a,m_a} L^{(5)}_{m_a}(z) 
 P_{k_a}^{1/2,1/2}(\cos2\phi_1) \right. \nb
 &&+ \left. \sum_{k_b,m_b} B_{k_b,m_b} 
L^{(5)}_{m_b}(z)P_{k_b}^{1/2,1/2}(\cos2\phi_3) \right\} \, , 
\label{psi00p} \\
&&\Psi_{(10)^-}(1,2) = 
\Xi_{0}(1,2) F_J e^{-z/2} 
\left\{ \sum_{k_a,m_a} A_{k_a,m_a} L^{(5)}_{m_a}(z) 
 \left[ \rho \sin{\phi_1} P_{k_a}^{3/2,1/2}(\cos2\phi_1) 
Y^M_1(\hat y_1) \right. \right. \nb
  &&+ \rho \cos{\phi_1}  \left. \left.  P_{k_a}^{3/2,1/2}(-\cos2\phi_1) 
Y^M_1(\hat x_1)\right] + \sum_{k_b,m_b} 
B_{k_b,m_b} L^{(5)}_{m_b}(z) \rho \sin{\phi_3} P_{k_b}^{3/2,1/2}(\cos2\phi_3) 
Y^M_1(\hat y_3)\right\} \, ,
\label{psi10m} \\
&&\Psi_{(11)^-}(1,2) = 
\Xi_{1}(1,2) F_J e^{-z/2} 
\left\{ \sum_{k_a,m_a} A_{k_a,m_a} L^{(5)}_{m_a}(z) 
 \left[ P_{k_a}^{1/2,3/2}(\cos2\phi_1) \rho \cos{\phi_1}
Y^M_1(\hat x_1) \right. \right. \nb
 &&-\rho \sin{\phi_1}  \left. \left. P_{k_a}^{1/2,3/2} (-\cos2\phi_1) 
Y^M_1(\hat y_1)\right] + \sum_{k_b,m_b} B_{k_b,m_b} L^{(5)}_{m_b}(z)
\rho \cos{\phi_3} P_{k_b}^{1/2,3/2}(\cos2\phi_3) Y^M_1(\hat x_3) \right\}  \, ,
\label{psi11m}
\eea
where $k_a\geq 0$ and $k_b\geq 1$. However, only even values of 
$k_a$ are allowed in $\Psi_{(00)^+}$.
In $^4$He$_2$-SF$_6$ we just study $\Psi_{(00)^+}$ and $\Psi_{(10)^-}$ 
states, since only  $S=0$ combinations are possible.

In Eqs.~(\ref{psi00p}--\ref{psi11m}) the linear 
coefficients, $\{A_{k_a,m_a}\}$ and $\{B_{k_b,m_b}\}$, are unknown 
quantities to be determined. 
The implementation of the variational principle for linear variational 
parameters leads to a generalized eigenvalues problem whose solutions,  
${E_i}$, are upper bounds to the true energy eigenvalues of the 
three-body Schr\"odinger equation. It is possible to improve the 
estimates of the energies by including a larger number of polynomials 
and channels in the CHH basis. If $N$ is the dimension of the basis set, 
the estimates ${E_i^N}$ will monotonically converge from above 
to the exact eigenvalues as $N$ is increased. The pattern of convergence 
 for the $(00)^+$ two lowest lying state of $^3$He$_2$-SF$_6$ is 
shown in Table~\ref{convergence}. 
An optimum choice of the non-linear parameter $\beta$ 
has been adopted to improve the convergence rate.

\section{Results}

Table~\ref{trimer_energies} collects the energies for the  
 He$_2$-SF$_6$ trimers in the uncorrelated, Jastrow and CHH approaches. 
The correlation function is set equal to unity in the uncorrelated 
calculations, whereas it has been choosen of the McMillan\cite{McMillan} 
form, 
\begin{equation}
f_J(r)= \exp \left[ 
-{1 \over 2} \left( {{b\sigma}\over r} \right)^5 \right]
 \, , 
\label{McMillan}
\end{equation}
in the variational case. Here, $\sigma$=2.556 \AA\ and $b$ is the 
only non-linear variational parameter. 
This type of correlation has been widely adopted 
in variational studies of liquid helium since it provides an excellent
 description of the short-range properties of the correlated wave 
function. In fact, it gives the exact short-range behavior for a 
12--6 Lennard--Jones atom--atom interaction. The correlation operator 
adopted in the CHH expansion has been described in the previous section.
In the CHH case we use different correlation functions  
since a non-linear parameter, $\beta$,  is already present
in the basis functions (see Eq.~\ref{chh}). 
Therefore, employing the McMillan form would imply a two 
non-linear parameters minimization. However, it has been
checked that the use of the McMillan correlation in the CHH expansion 
produces binding energies within $0.01\%$ of those given in 
Table~\ref{trimer_energies}, 
 showing that the converged results are to large extent independent of 
the correlation function, provided the short range behavior is adequately 
described.
  In Table~\ref{trimer_energies}, 
we also show the kinetic ($T$) and potential ($V$) contributions 
to the energy, separating the latter 
 in its dopant-helium (D-He) and helium-helium (He-He) parts. 

The uncorrelated trimers are unbound in all states, with the 
exception of the spatially antisymmetric (11)$^-$ for 
$^3$He$_2$-SF$_6$. The orbital antisymmetry reduces 
the probability of configurations having the two $^3$He atoms 
close to each other. Hence, the contribution of the strong 
He-He repulsion at short distances is drastically suppressed. 
In Ref.~\onlinecite{Jun01} an expansion of the single-particle helium 
$s$- and $p$-orbitals in a finite set of gaussian 
basis functions centered at the dopant provided an energy of 
-31.36 K  for the (11)$^-$ state in $^3$He$_2$-SF$_6$. 
This energy is higher than our uncorrelated estimate, pointing 
to a lack of convergence in the Hartree-Fock result in that reference. 

The introduction of the Jastrow correlation bounds all trimers,  
as it suppresses the short-range helium-helium repulsion. 
The L=0, positive parity states are the lowest lying ones, the 
other states having small excitation energies, lower than 1 K. 
The values of the variational parameter giving the minimum energies 
are also reported in the Table. The $b$-values for 
the spatially symmetric trimers are close to those found in 
the Jastrow correlated studies of bosonic liquid $^4$He. As in 
fermionic liquid $^3$He, $b$ is smaller for the spatially 
antisymmetric (11)$^-$ state, since both the correlation and the 
Pauli principle concur in depleting the $V_{\rm He-He}(r)$ repulsion. 

The converged CHH expansion provides slightly more binding 
(at most about -0.1 K) to the trimers. 
This fact is a strong indication of the high efficiency of 
the simple Jastrow correlated wave function in these systems. 
The kinetic and dopant-helium potential energies do not vary 
much in going from the uncorrelated to the variational and 
CHH estimates. The helium-helium potential energy is, instead, 
strongly dependent on the wave-function. For the uncorrelated 
cases the repulsive core is overwhelmingly dominant, except in 
the Pauli suppressed (11)$^-$ state. The short-range 
structure of the correlated and CHH wave functions results 
in a slightly attractive value of $\langle V_{\rm He-He}\rangle $ 
(from -0.3 K to -1.6 K).

For the mixed $^3$He-$^4$He-SF$_6$ trimer we give only the 
energies of the lowest lying (0)$^+$ state. The extension of the 
CHH theory presented in Section III to this type of trimer is 
straightforward.  All energies consistently sit in between 
the lighter  $^3$He$_2$-SF$_6$ and the heavier 
$^4$He$_2$-SF$_6$ cases.
  
It is worth noticing that the strong supression of the mutual He-He 
repulsion due to the correlations translates into a total binding 
energy which is very close to the sum of the two dimer energies. 
The sum of the $\ell = 0$ energies
of Table~\ref{single_obser} gives $-54.67$, $-61.13$ and $-57.90$~K,
respectively, for the combinations $^3$He$_2$, $^4$He$_2$ and $^3$He-$^4$He,
which are very close to the binding energies of the trimer. The practical
effect of the correlations is to reduce the He-He interaction to a
small attraction of $\approx 0.1-0.2$~K. This result is very different
from the findings of Ref.~\onlinecite{Jun01}, with total binding energies
smaller than ours by roughly a factor of two. 

We use a simple argument to estimate the accuracy of the infinite dopant mass 
approximation.  From Table III we observe that, to a very 
good extent, the He$_2$-SF$_6$ trimer can be considered as the 
superposition of two independent He-SF$_6$ dimers. Accordingly,
the $\alpha$--$\beta$--trimer corrected kinetic energy results
$T^{(\alpha,\beta)}=T^{(\alpha,\beta)}_{M=\infty}
[1+1/2(m_\alpha/M+m_\beta/M)]$. 
This correction corresponds to a modification of the total trimer 
energy less than $1\%$.

Structural properties of the trimers are shown in 
Table~\ref{trimer_observables}. We give the root-mean-square 
(rms) dopant-helium distance, $\sqrt {\langle R^2 \rangle }$, the 
rms helium-helium distance, $\sqrt {\langle r^2_{12} \rangle }$, 
and the average value of the cosine between the two D-He 
radii, $\langle \cos (\theta_{12}) \rangle$, in the three 
approaches. $\sqrt {\langle R^2 \rangle }$ is not very 
sensitive to the introduction of the He-He correlation. 
For a given trimer, it assumes essentially the same values   
in the different states, reflecting the fact that the 
D-He wave functions, $\phi_{1s}$ and $\phi_{1p}$, are  
similar. 
Since the He-SF$_6$ interaction is the same for both helium
isotopes, the smaller $^3$He mass would  produce a larger
kinetic energy than $^4$He. In order to minimize the energy, 
this tendency is partially compensated by a larger value of 
  $\sqrt {\langle R^2_{{\rm D}-^3{\rm He}} \rangle }$ 
with respect to $\sqrt {\langle R^2_{{\rm D}-^4{\rm He}} \rangle }$.
The He-He average distance increases 
in going form the uncorrelated to the variational and CHH cases 
in the spatially symmetric states, as a consequence of the 
intoduction of the correlation, which suppresses short-range 
He-He configurations. The effect is not very visible in the 
spatially antisymmetric state, (11)$^-$, being these 
configurations already largely inhibited by the Pauli 
principle. We even find a small decrease of 
$\sqrt {\langle r^2_{12} \rangle }$ in this state after solving 
the CHH equations. The values of the uncorrelated average He-He 
cosine, $\langle \cos (\theta_{12}) \rangle$, are immediately 
understood in terms of the structure of $\Phi_{(LS)^\pi}(1,2)$.  
Correlations change these values, mostly in the spatially 
symmetric $S=0$ states. For instance, the average 
cosine corresponds to $\theta_{12}=0.39 \, \pi $ in the 
uncorrelated $(10)^-$ state of the $^3$He$_2$-SF$_6$ trimer, 
and to $\theta_{12}=0.45 (0.44) \, \pi $ for the Jastrow (CHH) 
case. The change for the spatially antisymmetric state, 
$(11)^-$, is much less evident.

\begin{figure}
 \includegraphics*[width=12cm]{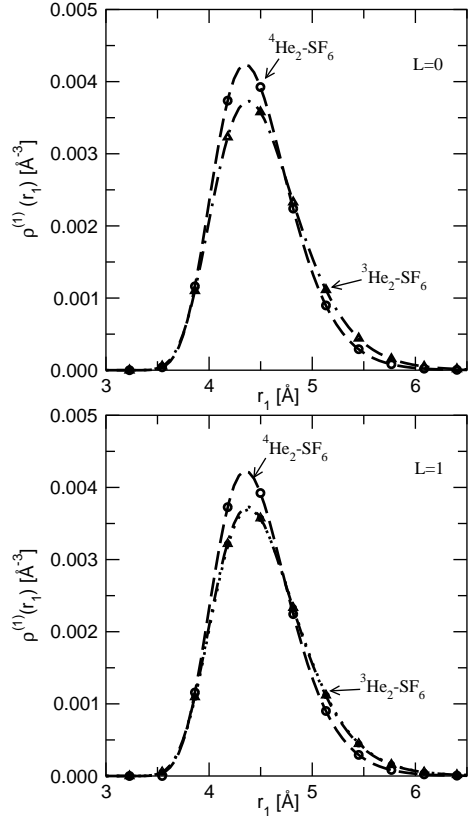}
\caption{ Helium densities in the $L$=0 (upper panel) and 
$L$=1 (lower panel) states of the He$_2$-SF$_6$ trimers. The symbols 
denote the uncorrelated densities, the lines stand for the CHH ones. 
In the upper panel the circles and the dashed line stand for 
$^4$He$_2$-SF$_6$ while the triangles and the dash-dotted line correspond to
$^3$He$_2$-SF$_6$. The same notation is used for the L=1 $^4$He$_2$-SF$_6$ case 
in the lower panel. For the L=1 $^3$He$_2$-SF$_6$ trimer two states are reported, 
corresponding to S=0 (triangles and dotted line) and S=1 (stars and dash-dotted line).  
However, at the scale of the Figure they are hardly distinguishible.}
\label{fig:OBD}
\end{figure}

The one-body helium densities (OBD), 
\begin{equation}
\rho^{(1)}_{\gamma}({\bf r}_1)= {
{\int d{\bf r}_2 \vert \Psi_{\gamma}(1,2) \vert ^2 }
\over
{\int d{\bf r}_1\int d{\bf r}_2 \vert \Psi_{\gamma}(1,2) \vert ^2 }
} 
\label{OBD}
\end{equation}
normalized as
\begin{equation}
\int d{\bf r}_1 \, \rho^{(1)}_{\gamma}({\bf r}_1)\,=\, 1 \, , 
\label{OBD_norm}
\end{equation}
are shown in Figure~\ref{fig:OBD} for the $L$=0 and 1 states 
of the $^3$He$_2$-SF$_6$ and  $^4$He$_2$-SF$_6$ trimers, obtained 
in the uncorrelated (symbols) and CHH (lines) approaches. 
The OBDs are very little affected by both the introduction of the He-He 
correlation and by the optimization of the D-He wave function. 
Actually, they are similar to the dimer radial probability densities 
shown in Fig.~\ref{probdens}. As in the dimer case, the $^4$He atom 
is more localized than the $^3$He one because of its 
larger mass. For a given trimer, the OBDs  
do not appreciably depend on the $L$-values, since the $1s$ and 
$1p$ D-He wave functions are almost coincident.

\begin{figure}
 \includegraphics*[width=12cm]{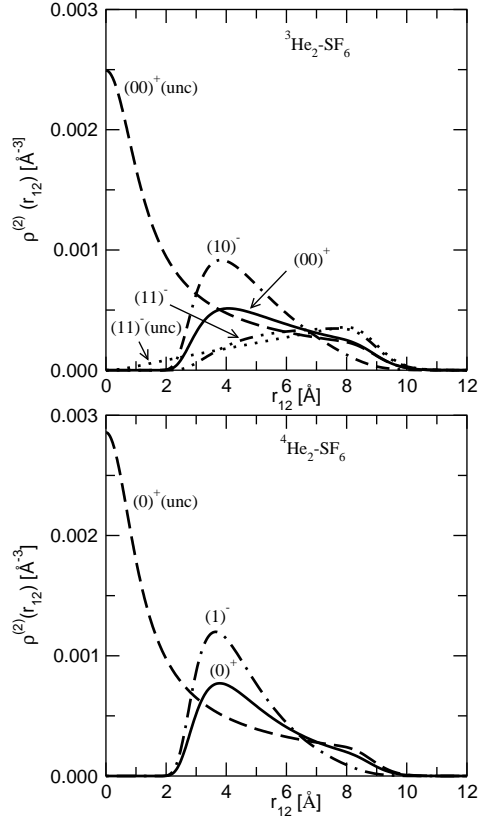}
\caption{ Helium-helium center-of-mass integrated two-body densities 
in the $^3$He$_2$-SF$_6$ (upper panel) and $^4$He$_2$-SF$_6$ 
(lower panel) trimers in the uncorrelated and CHH approaches. The uncorrelated
(10)$^-$ two body density for $^3$He$_2$-SF$_6$ 
is very close to the (00)$^+$ one and it is not shown in the Figure. 
The same holds for the uncorrelated (1)$^-$ TBD for $^4$He$_2$-SF$_6$.} 
\label{fig:TBD}
\end{figure}

Differences between the various approaches show up in the 
helium-helium two-body density (TBD), defined as:
\begin{equation}
\rho^{(2)}_{\gamma}({\bf r}_1,{\bf r}_2)= {
{\vert \Psi_{\gamma}(1,2) \vert ^2 }
\over
{\int d{\bf r}_1\int d{\bf r}_2 \vert \Psi_{\gamma}(1,2) \vert ^2 }
} \, \, . 
\label{TBD}
\end{equation}
In Figure~\ref{fig:TBD} we display the center-of-mass integrated 
TBD, 
\begin{equation}
\rho^{(2)}_{\gamma}({\bf r}_{12})= 
\int d{\bf R}_{12} \, \rho^{(2)}_{\gamma}({\bf r}_1,{\bf r}_2) 
\label{TBD_12}
\end{equation}
normalized as
\begin{equation}
\int d{\bf r}_{12} \, \rho^{(2)}_{\gamma}({\bf r}_{12})
\,=\, 1 \, , 
\end{equation}
where ${\bf R}_{12}=(m_1 {\bf r}_1+m_2 {\bf r}_2)/(m_1+m_2)$ 
is the He-He center-of-mass coordinate and ${\bf r}_{12}={\bf r}_1-{\bf r}_2$ 
is the He-He distance.  $\rho^{(2)}_{\gamma}({\bf r}_{12})$ 
gives the probability of the two helium atoms being at a 
distance $r_{12}$ apart.

The Pauli repulsion suppresses $\rho^{(2)}_{\gamma}({\bf r}_{12})$ in 
the uncorrelated (11)$^-$ state of the $^3$He-SF$_6$ trimer at short 
He-He distances, in contrast with the uncorrelated (00)$^+$ one. 
This behavior makes the former state bound and the latter unbound. 
We recall that the repulsive core of the He-He interaction is 
$R_c\sim \, 2.5$  \AA. The introduction of the He-He correlation 
depletes the TBD at small $r$-values in all the states, which result, 
as a consequence, all bound. 
In both trimers the helium atoms are more closely packed in the 
spatially symmetric, $L=1$ states, consistently with the values of 
$\sqrt {\langle r^2_{12} \rangle }$ shown in 
Table~\ref{trimer_observables}. As expected, the spatially 
antisymmetric (11)$^-$ state is the most diffuse.
The uncorrelated long-range structures of the TBDs remain   
essentialy untouched by  the correlations.

\begin{figure}
 \includegraphics*[width=15cm]{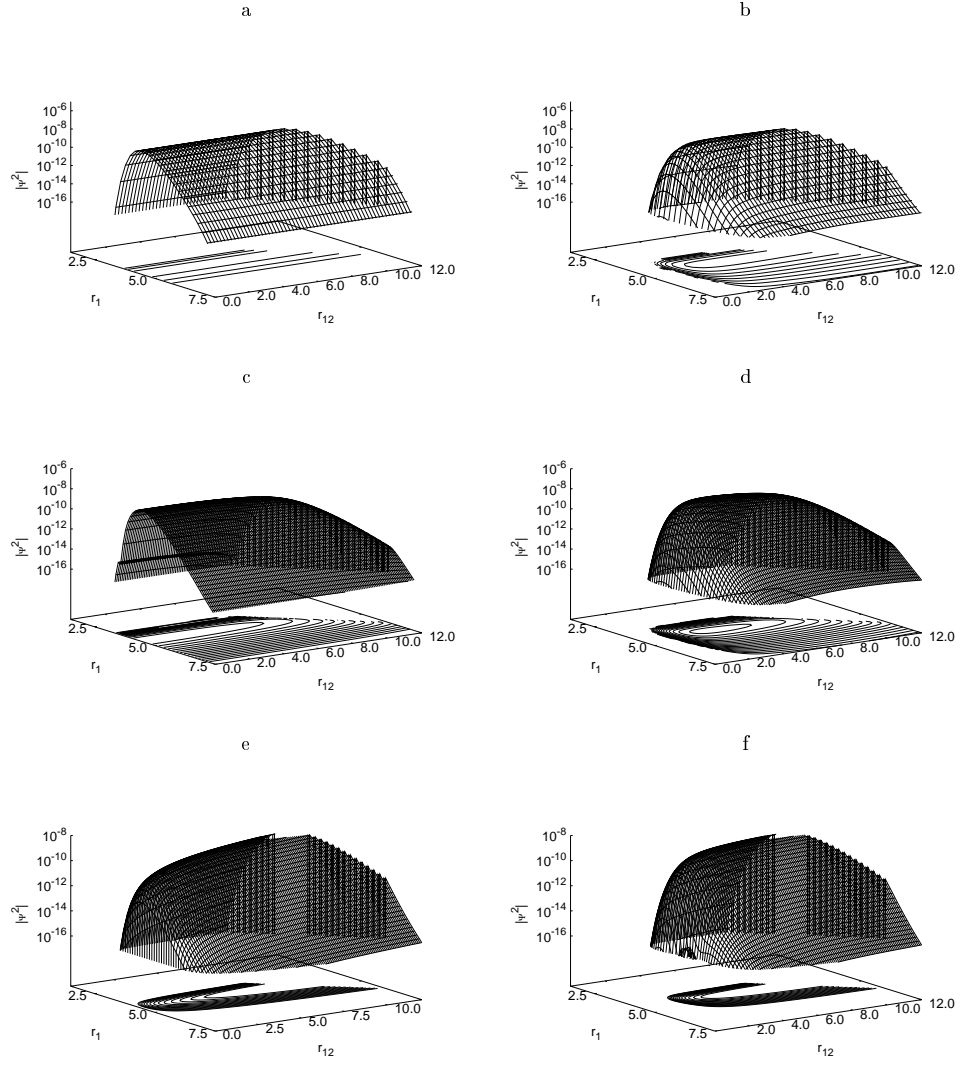}
\caption{ Helium-helium two-body densities in the isosceles 
configurations,  $\rho^{(2)}_{\gamma}(r_1,r_2=r_1,r_{12})$, for  
the $^3$He$_2$-SF$_6$ trimer in the uncorrelated (00)$^+$ ($a$), 
(10)$^-$ ($c$) and (11)$^-$ ($e$) states, and in 
the CHH (00)$^+$ ($b$), (10)$^-$ ($d$) and (11)$^-$ ($f$) states. 
Distances in \AA \, and densities in \AA$^{-6}$.}
\label{fig:TBD_3D}
\end{figure}

We finally show in Figure~\ref{fig:TBD_3D} the TBD around 
the SF$_6$ molecule in the isosceles configurations, 
$\rho^{(2)}_{\gamma}(r_1,r_2=r_1,r_{12})$, for the 
uncorrelated and CHH (00)$^+$, (10)$^-$ and 
(11)$^-$ states of $^3$He$_2$-SF$_6$. All of the TBDs 
vanish at low $r_1$ values because of the strong D-He 
repulsion. The isotropic distribution shown by the 
uncorrelated (00)$^+$ TBD disappears after introducing 
the He-He correlation, which suppresses the density at low inter-helium 
distances. As already noticed in Fig.~\ref{fig:TBD}, the 
(11)$^-$ TBD is the least affected by the correlations since 
it displays a short range He-He repulsion due to the Pauli 
principle.

\section{Summary and conclusions}

The study of He$_2$-SF$_6$ trimers reveals the crucial role of the dopant 
heavy molecule in binding these systems. In fact, the SF$_6$ molecule 
acts as a fixed center of force in which the He atoms are moving. We have shown 
that the specific features of the SF$_6$-He interaction gives a rotational 
band in the excitation spectrum of the dimers. The different masses of the 
$^3$He and $^4$He explain in a qualitative way the particular features of each 
dimer. The solution of the dimer Schr\"odinger equation provides the 
single-particle 
 wave functions needed to build the trial wave function to be used in the 
variational study of the trimers. 

The ground state energies of the three isotopic trimers, namely 
$^4$He$_2$-SF$_6$, $^3$He$_2$-SF$_6$ and $^4$He-$^3$He-SF$_6$, have been 
estimated employing a Jastrow correlated wave-function built up as the 
product of the dimer wave functions times a two-body correlation function 
of the McMillan type between the He atoms, having a single variational 
 parameter. The accuracy of the variational approach has been tested 
against the Correlated Hyperspherical Harmonics expansion method. 
We have found that the variational results are in excellent agreement with 
the CHH ones at convergence. 

The role of the Jastrow correlation is crucial in order to overcome the 
strong repulsion between the He atoms. 
Actually the uncorrelated variational approach does not bind the trimers, 
except the $^3$He$_2$-SF$_6$ one in the (11)$^-$ configuration. 
The reason is that its wave function is spatially antisymmetric, 
and therefore the two $^3$He atoms are kept already 
apart by the Pauli repulsion. 
We stress that the preferred spatial configuration 
assumed by the two He atoms is such that they take advantage from the mutual 
 attraction, suppressing, as much as possible, the short range repulsion. 
As a result, the optimal configuration is not a linear one, with the SF$_6$ 
 molecule in the middle of the two He atoms. 
Instead, the He atoms are closer, and their
position vectors with respect to the SF$_6$ molecule form an angle
between $\simeq 70^{\rm o}$ and $\simeq 86^{\rm o}$, depending on the
particular state. The practical effect of the correlations is that the binding 
energy of the trimer is slighlty larger than the sum of the binging energies 
of the corresponding dimers. 

The He-He correlation does not particularly affect the 
the helium probability densities in the trimers, which are 
 similar to those in the dimer. In contrast, the correlation 
is essential in inverting the energy hierarchy between the 
spatially symmetric and antisymmetric configurations. In fact,   
in the correlated $^3$He$_2$-SF$_6$  trimer the (00)$^+$ state, 
symmetric in space with both He atoms in the $1s$ state and $S=0$, 
is more bound than the (11)$^-$ state, antisymmetric in space with 
one atom in $1s$ and the other in the $1p$ state and $S=1$ (aligned spins). 
The uncorrelated approach does not even bind the (00)$^+$ trimer, whereas 
the (11)$^-$ one is still bound. 

The good agreement between the variational and the CHH estimates 
 for the trimers makes us confident that medium size He doped clusters 
can be accurately described by means of a correlated variational  
wave function. This trial wave function would be built up from the 
single-particle wave functions obtained after solving the dimer case, 
and from an appropriate Jastrow factor to properly take into account 
He-He correlations. In this respect, Variational Monte Carlo  and 
Fermi Hypernetted Chain techniques seem to be the most likely candidates 
to microscopically address the study of medium-heavy doped helium nanodroplets.

\section*{Acknowledgments}

Fruitful discussions with Stefano Fantoni and Kevin Schmidt 
are gratefully acknowledged. This work has been partially 
supported by DGI (Spain) grants BFM2001-0262, BFF2002-01868,
Generalitat Valenciana grant GV01-216, 
Generalitat de Catalunya project 2001SGR00064, and by the Italian
MIUR through the PRIN: {\it Fisica Teorica del Nucleo Atomico e dei
Sistemi a Molti Corpi}.  



\pagebreak

\begin{table}
\begin{center}
\begin{tabular}{c|ccccc|ccccc}
& \multicolumn{5}{c|}{$^3$He} & \multicolumn{5}{c}{$^4$He} \\ \hline  
$\ell$ & $e_{0\ell}$ & $\langle V \rangle$ &
$\langle R \rangle$ & $\sqrt{\langle R^2 \rangle}$ & 
$e_{1\ell}$ & $e_{0\ell}$ & $\langle V \rangle$ &
$\langle R \rangle$ & $\sqrt{\langle R^2 \rangle}$ & 
$e_{1\ell}$ \\ \hline
0 & -27.336 &  -39.184 & 4.622 & 4.646 & -1.940 &
    -30.563 &  -41.509 & 4.549 & 4.568 & -4.292 \\
1 & -26.561 &  -39.104 & 4.626 & 4.650 & -1.541 &
    -29.964 &  -41.459 & 4.552 & 4.571 & -3.920 \\
2 & -25.016 &  -38.939 & 4.635 & 4.659 & -0.775 &
    -28.767 &  -41.355 & 4.557 & 4.577 & -3.184 \\
3 & -22.707 &  -38.677 & 4.648 & 4.673 &  &
    -26.976 &  -41.193 & 4.566 & 4.585 & -2.105 \\
4 & -19.647 &  -38.298 & 4.667 & 4.693 &  &
    -24.598 &  -40.963 & 4.577 & 4.597 & -0.721 \\
5 & -15.854 &  -37.771 & 4.693 & 4.720 &  &
    -21.641 &  -40.653 & 4.592 & 4.613 &  \\
6 & -11.354 &  -37.047 & 4.728 & 4.758 &  &
    -18.117 &  -40.243 & 4.612 & 4.634 &  \\
7 &  -6.183 &  -36.041 & 4.776 & 4.809 &  &
    -14.040 &  -39.707 & 4.637 & 4.660 &  \\
8 &  -0.394 &  -34.579 & 4.848 & 4.887 &  &
     -9.431 &  -39.001 & 4.669 & 4.693 &  \\
9 &         &          &       &       &  &
     -4.318 &  -38.058 & 4.711 & 4.738 &  \\ \hline
\end{tabular}
\end{center}
\caption{Observables of the bound states of the  He-SF$_6$ dimers.
Energies are in K and distances in \AA.}
\label{single_obser}
\end{table}

 \begin{center}
 \begin{table}[h]
 \begin{tabular}{c|cccc|ccc} 
  & \multicolumn{4}{c|} {$(00)^+$ ground state} & \multicolumn{3}
{c}{$(00)^+$ first excited state} \\
\hline
 $n_3 \backslash n_1$ & 1 & 2 & 3 & 4 & 1 & 2 & 3 \\
\hline
  0 & -54.269 & -54.652 & -54.656 & -54.656 & - & - & - \\
  1 & -54.381 & -54.743 & -54.746 & -54.746 & -51.652 & -51.674 & -51.674 \\
  2 & -54.397 & -54.777 & -54.780 & -54.780 & -52.384 & -52.416 & -52.416 \\
  3 & -54.413 & -54.789 & -54.792 & -54.792 & -52.411 & -52.439 & -52.439 \\
  4 & -54.414 & -54.792 & -54.795 & -54.795 & -52.425 & -52.454 & -52.454 \\
  5 & -54.416 & -54.793 & -54.796 & -54.796 & -52.426 & -52.455 & -52.455 \\
  6 & -54.416 & -54.793 & -54.796 & -54.796 & -52.427 & -52.456 & -52.456 \\
  7 & -54.417 & -54.793 & -54.796 & -54.797 & -52.427 & -52.457 & -52.457 \\
  8 & -54.417 & -54.793 & -54.796 & -54.797 & -52.427 & -52.457 & -52.457 \\
\hline
 \end{tabular}
 \caption{Convergence pattern of $E_i^{N}$ for 
the two lowest lying states of the $(00)^+$ band in $^3$He$_2$-SF$_6$ 
 as a function of the numbers ($n_1,n_3$) of Jacobi polynomials in 
the expansion (\ref{psi00p}) for the amplitude (1,2) and 3, respectively. 
The number of Laguerre polynomials is independently optimized for each case. 
The variational parameter $\beta$ is fixed to 
$1.59$ \AA. Energies in K.} 
\label{convergence}
 \end{table}
 \end{center}


\begin{table}[h!]
\begin{center}
\begin{tabular}{c|ccc|cc|c} 
& \multicolumn{3}{c|}{$^3$He-$^3$He} 
& \multicolumn{2}{c|}{$^4$He-$^4$He} 
& \multicolumn{1}{c}{$^3$He-$^4$He} \\ 
\hline  
 & (00)$^+$ & (10)$^-$& (11)$^-$
 &  (0)$^+$ & (1)$^-$ & (0)$^+$ \\ \hline
 $b$ & 1.16 & 1.17 & 1.06 & 1.16 & 1.17 & 1.16 \\
 $E^{(unc)}$ & 837.18 & 1704.7 & -33.378 & 931.84 & 1899.2 & 881.18 \\
 $E^{(var)}$ & -54.746 & -53.859 & -54.078 & -61.346 & -60.783 & -58.045 \\
 $E^{(CHH)}$ & -54.797 & -53.871 & -54.161 & -61.442 & -60.843 & -58.102  \\
 $T^{(unc)}$ & 23.696 & 24.390 & 24.390 & 21.892 & 22.440 & 22.794 \\
 $T^{(var)}$ & 23.936 & 24.948 & 24.389 & 22.074 & 22.869 & 23.004 \\
 $T^{(CHH)}$ & 24.301 & 25.587 & 24.553 & 22.668 & 23.713 & 23.528 \\
 $V^{(unc)}_{D-He}$ & -78.368 & -78.288 & -78.288 & -83.018 & -82.967 & -80.693 \\
 $V^{(var)}_{D-He}$ & -78.045 & -77.661 & -78.213 & -82.761 & -82.469 & -80.401 \\
 $V^{(CHH)}_{D-He}$ & -78.392 & -78.252 & -78.402 & -83.058 & -82.988 & -80.702 \\
 $V^{(unc)}_{He-He}$ & 891.85 & 1758.6 & 20.520 & 992.97 & 1959.7 & 939.08 \\
 $V^{(var)}_{He-He}$ & -0.638 & -1.146 & -0.254 & -0.660 & -1.183 & -0.649 \\
 $V^{(CHH)}_{He-He}$ & -0.705 & -1.206 & -0.312 & -1.052 & -1.569 &  -0.908 \\
\hline
\end{tabular}
\end{center}
\caption{Total, kinetic and potential energies (in K) of the He$_2$-SF$_6$ 
trimers in the uncorrelated, variational and CHH approaches. 
$b$ is the adimensional parameter of the McMillan correlation function. }
\label{trimer_energies}
\end{table}
 

\begin{table}[h!]
\begin{center}
\begin{tabular}{c|ccc|cc|c} 
& \multicolumn{3}{c|}{$^3$He-$^3$He} 
& \multicolumn{2}{c|}{$^4$He-$^4$He} 
& \multicolumn{1}{c}{$^3$He-$^4$He} \\ 
\hline  
 & (00)$^+$ & (10)$^-$& (11)$^-$
 &  (0)$^+$ & (1)$^-$ & (0)$^+$ \\ \hline
$\sqrt{\langle R^2 \rangle}^{(unc)}$& 4.646 & 4.648 &  4.648 & 4.568 & 4.570 & 4.646/4.568 \\ 
$\sqrt{\langle R^2 \rangle}^{(var)}$& 4.655 & 4.665 & 4.650 & 4.576 & 4.584 & 4.656/4.575 \\
 $\sqrt{\langle R^2 \rangle}^{(CHH)}$& 4.645 & 4.645 & 4.653 & 4.568 & 4.569 & 4.646/4.567 \\
$\sqrt{\langle r_{12}^2 \rangle}^{(unc)}$& 6.570 & 5.382 & 7.580 & 6.461 & 5.288 & 6.516 \\
 $\sqrt{\langle r_{12}^2 \rangle}^{(var)}$& 6.996 & 6.029 & 7.642 & 6.893 & 5.946 & 6.944 \\
$\sqrt{\langle r_{12}^2 \rangle}^{(CHH)}$& 6.845 & 5.903 & 7.476 & 6.351 & 5.363 & 6.533 \\
$\langle \cos(\theta_{12})\rangle^{(unc)}$ & 0 & 1/3 & -1/3 & 0 & 1/3 & 0 \\
 $\langle \cos(\theta_{12})\rangle^{(var)}$ & -0.133 &  0.163 & -0.355 & -0.138 & 0.157 & -0.135 \\
 $\langle \cos(\theta_{12})\rangle^{(CHH)}$ & -0.087 & 0.195 & -0.298 & 0.033 & 0.260 & -0.006  \\
\hline
\end{tabular}
\end{center}
\caption{Root mean square dopant-helium, $\sqrt{\langle R^2 \rangle}$, and 
helium-helium, $\sqrt{\langle r_{12}^2 \rangle}$, radii and 
average cosine between the two dopant-helium radii, 
$\langle \cos(\theta_{12} \rangle$ in the uncorrelated, 
variational and CHH approaches. 
In the first two rows of the last column, the first (second) value 
gives the D-$^3$He (D-$^4$He) rms radius. 
Distances in \AA.}
\label{trimer_observables}
\end{table}

\pagebreak
 
\end{document}